\documentclass[aps,prl, preprintnumbers,amsmath,amssymb,latexsym,array,enumerate,letter,twocolumn,superscriptaddress]{revtex4-2}

\usepackage{amssymb}
\usepackage{amsmath}
\usepackage{epsfig}
\usepackage{hyperref}
\usepackage{breakurl}
\usepackage{xcolor}
\usepackage{slashed}
\usepackage{dsfont,bbm}
\usepackage{mathrsfs}

\usepackage{multirow, graphicx, subfigure, placeins, float}

\usepackage{extarrows}
\makeatletter
\def\xrightleftharpoonsfill@{%
  \arrowfill@\leftharpoondown\relbar\rightharpoonup}
\providecommand{\xrightleftharpoons}[2][]{%
  \ext@arrow 0099\xrightleftharpoonsfill@{#1}{#2}}
\makeatother

\usepackage{enumitem,kantlipsum}

\usepackage{todonotes}
\usepackage{tikz}
\usetikzlibrary{arrows.meta,positioning,calc}
\usepackage[compat=1.1.0]{tikz-feynman}
\tikzfeynmanset{warn luatex=false}

\graphicspath{{./fig/}}

\newcommand{\tr}{\text{tr}}
\newcommand{\ff}{\boldsymbol{F}}
\newcommand{\cf}{\mathcal{F}}
\newcommand{\amp}{\boldsymbol{A}}
\newcommand{\ca}{\mathcal{A}}
\newcommand{\gA}{\boldsymbol{M}}
\newcommand{\La}{\mathcal{L}}
\newcommand{\LaYMS}{\La^{\text{YMs}}}
\newcommand{\klt}{\mathbf{S}}
\newcommand{\clsA}{\mathrm{I}}
\newcommand{\blc}{\mathcal{B}}

\newcommand{\ampnh}[2]{\amp_{#2}^{[#1]}}
\newcommand{\campnh}[2]{\ca_{#2}^{[#1]}}
\newcommand{\ganh}[2]{\gA_{#2}^{[#1]}}
\newcommand{\cffnh}[2]{\cf_{#2}^{[#1]}}

\newcommand{\clrBlue}{\color{cyan!70!black}}
\newcommand{\clrOrange}{\color{orange!87!black}}
\definecolor{letterGreen}{RGB}{73,175,27}
\newcommand{\clrLetterGreen}{\color{letterGreen}}
\newcommand{\bH}{\bar{H}}

\newcommand{\Ca}{\amp}
\newcommand{\Cf}{\ff}
\newcommand{\fftabc}{\ff_3(1^\phi, 2^\phi, 3^g)}
\newcommand{\cftracb}{\cf_3(1^\phi, 3^g, 2^\phi)}
\newcommand{\ampppgq}{\amp_{4}(1^\phi, 2^\phi, 3^g, q^{\bH})}
\newcommand{\gapphq}{\gA_{4}(1^\phi, 2^\phi, 3^h, q^{H})}
\newcommand{\catracbq}{\ca_4(1^\phi, 3^g, 2^\phi, q^{\bH})}
\newcommand{\catrcabq}{\ca_4(3^g, 1^\phi, 2^\phi, q^{\bH})}

\newcommand{\dampPPGHH}{\ampnh{2}{5}({1^\phi, 2^\phi, 3^g, q_1^{\bH_1}, q_2^{\bH_2}})}
\newcommand{\trsA}{[12q_2q_1]}
\newcommand{\ampTrsA}{\mathscr{A}^{\trsA}_{5}}
\newcommand{\ppmAppghhI}{\Theta_{3 \times 3}^{{\trsA}}}

\begin{document}

\preprint{USTC-ICTS/PCFT-26-35}
\title{Double copy of form factors with multiple operator insertions}
\author{Xinyue Li}
\email{lixinyue251@mails.ucas.ac.cn}
\affiliation{School of Fundamental Physics and Mathematical Sciences, Hangzhou Institute for Advanced Study, UCAS, Hangzhou 310024, China}
\affiliation{Institute of Theoretical Physics, Chinese Academy of Sciences, Beijing 100190, China}
\affiliation{School of Physical Sciences, University of Chinese Academy of Sciences, Beijing 100049, China}
\author{Gang Yang}
\email{yangg@itp.ac.cn}
\affiliation{School of Fundamental Physics and Mathematical Sciences, Hangzhou Institute for Advanced Study, UCAS, Hangzhou 310024, China}
\affiliation{Institute of Theoretical Physics, Chinese Academy of Sciences, Beijing 100190, China}
\affiliation{School of Physical Sciences, University of Chinese Academy of Sciences, Beijing 100049, China}
\affiliation{Peng Huanwu Center for Fundamental Theory, Hefei, Anhui 230026, China}
\author{Guorui Zhu}
\email{zhuguorui@itp.ac.cn}
\affiliation{Institute of Theoretical Physics, Chinese Academy of Sciences, Beijing 100190, China}
\affiliation{School of Physical Sciences, University of Chinese Academy of Sciences, Beijing 100049, China}

\begin{abstract}
Extending the double copy of scattering amplitudes to more general physical quantities involving local gauge-invariant operators is a central open question. While progress has been made in the double copy of form factors (FFs) with a single-operator insertion, it has led to two intriguing features:
poles that are spurious from the FF viewpoint become physical propagators in gravity, and FFs obey hidden factorization relations on these poles.
This picture is difficult to generalize to FFs with multiple operator insertions due to even more complicated spurious pole structures.
We resolve this problem by introducing a ``dyeing'' procedure that promotes color-singlet operators to adjoint colored states,
while the original FF is recovered by the inverse ``bleaching'' (U(1) decoupling) operation.
In this new picture, the spurious poles are propagators of the dyed state, and the hidden factorization relations follow from BCJ relations of the dyed amplitudes.
For multiple operator insertions, this framework uncovers a new scalar-ordering structure that survives the double copy to gravity.
\end{abstract}

\maketitle

\section{Introduction}

\noindent
The double copy relates gauge and gravity scattering amplitudes through
Kawai--Lewellen--Tye (KLT) relations \cite{Kawai:1985xq}, Bern--Carrasco--Johansson (BCJ)
color-kinematics duality \cite{Bern:2008qj,Bern:2010ue}, and the Cachazo--He--Yuan (CHY) formula
\cite{Cachazo:2013hca,Cachazo:2014xea}
(see~\cite{Bern:2019prr,Bern:2022wqg,Adamo:2022dcm} for reviews).
It is natural to ask whether this relation can be extended to observables
involving local gauge-invariant operators.

A prominent class of such quantities consists of form factors (FFs),
defined as matrix elements between $n$ on-shell states and a gauge-invariant operator $\mathcal{O}$
\cite{Maldacena:2010kp,Brandhuber:2010ad,Bork:2010wf},
\begin{equation}
\ff_{\mathcal{O},n} = \int d^{D} x\, e^{i q \cdot x}
\langle 1 \cdots n |\mathcal{O}(x)| 0\rangle \,.
\end{equation}
Previous work \cite{Lin:2021pne,Lin:2022jrp,Lin:2023rwe} developed the double copy of tree-level FFs and revealed two intriguing features.
First, FF numerators satisfying color-kinematics duality contain poles that are spurious in the gauge-theory FF
but transmute to physical propagators after the double copy.
Second, FFs obey hidden factorization relations at these poles.
The physical origin of these features remains unexplained.
Furthermore, it is highly desirable to generalize the double copy to FFs with multiple operator insertions (often referred to as generalized FFs) ~\cite{Engelund:2012re,Gao:2013dza,Koster:2016fna, Ahmed:2019yjt}, which arise naturally in on-shell approaches to correlation functions.
However, the construction in \cite{Lin:2021pne,Lin:2022jrp,Lin:2023rwe} is difficult to extend to such cases, since more complicated spurious pole structures prevent a straightforward, consistent formulation.

In this Letter we propose a new perspective on the FF double copy that helps resolve these problems.
The starting point is to represent the operator insertion using an auxiliary
color-singlet massive Higgs scalar $H$, in analogy to the Higgs effective theory \cite{Wilczek:1977zn, Shifman:1979eb, Dawson:1990zj, Kniehl:1995tn, Chetyrkin:1997un}, where the FF is mapped to an amplitude with an extra color-singlet particle.
We then \emph{dye} the singlet scalar into an adjoint massive particle:
\begin{equation}
  H \quad \xrightleftharpoons[\text{bleach}]{\text{dye}} \quad \bar H = H^a T^a \,,
\end{equation}
which embeds the FF into a fully colored amplitude---the dyed amplitude.
As detailed later, this procedure introduces symmetric color tensors into the interaction vertices.
The physical FF can be recovered via the inverse \emph{bleaching} operation,
which replaces the color generator of the dyed leg with the identity matrix --- a step equivalent to the $U(1)$ decoupling operation.

Within this framework, the two aforementioned features admit a natural physical interpretation.
The spurious poles are simply standard propagators of the dyed particle; they
are invisible in the FF because bleaching projects out their color
factors, but their kinematic counterparts survive the double copy.
Furthermore, the hidden factorization relations are directly inherited from the BCJ relations of the dyed amplitude by taking the corresponding massive-channel on-shell limits.

Importantly, this framework generalizes naturally to FFs with multiple operator insertions.
Each operator insertion is mapped to a dyed Higgs leg, lifting the multi-operator FF to a multi-Higgs amplitude.
Through this construction, we uncover a novel structure: \emph{scalar orderings}.
The resulting amplitudes decompose into gauge-invariant sectors labeled by
these orderings, characterized by  a block-diagonal propagator matrix.
This scalar-ordering decomposition survives the double copy, yielding
independent, scalar-ordered gravity sectors.

\section{Dyeing and bleaching}

\noindent
As the primary examples, we use the FFs of $\mathcal O=\tr(\phi^2)$ in the Yang--Mills-scalar theory,
with Lagrangian
\begin{equation}
  \LaYMS =
  -\frac{1}{4}F_{\mu\nu}^aF^{a,\mu\nu}
  +\frac{1}{2}(D_\mu\phi)^a(D^\mu\phi)^a \, .
\end{equation}
Here $\phi=\phi^aT^a$ is a scalar field in the adjoint representation of
$SU(N)$, with $T^a$ the generators of the group.

The FF can be viewed as an amplitude with one additional Higgs leg with $q^2=m^2$:
\begin{equation}
  \ff_{\tr(\phi^2),n}(\{p_i\})
  =
  \amp_{n+1}(\{p_i\},q^H) ,
  \label{eq:ff_as_singlet_amp}
\end{equation}
where the operator insertion is interpreted as an effective vertex $H\tr(\phi^2)$.
Since $H$ is a color singlet, it carries no color charge and has no
gauge coupling along the $H$ line.

The dyeing operation promotes the color-singlet $H$ to an adjoint massive
state $\bar H = H^a T^a$.
We refer to this state as the dyed Higgs.
The Higgs kinetic term and the effective interaction vertex are dyed as
\begin{gather}
  \frac{1}{2}(\partial_\mu H)(\partial^\mu H)
  \xrightarrow{\ \text{dye}\ }
  \frac{1}{2}(D_\mu H)^a(D^\mu H)^a,
  \label{eq:dyed_kinetic_terms}
  \\
  H\tr(\phi^2)
  \xrightarrow{\ \text{dye}\ }
  d^{abc} H^a \phi^b \phi^c .
  \label{eq:dyed_operator_vertex}
\end{gather}
These two changes have distinct roles.
(i) The covariant derivative gives $\bar H$ the standard gauge couplings
of an adjoint scalar, including $g\bar H\bar H$ and $gg\bar H\bar H$
vertices, and hence introduces additional Feynman graphs.
(ii) The operator vertex is colored by the symmetric tensor
$d^{abc}=\tr(T^a\{T^b,T^c\})$, because $\tr(\phi^2)$ is symmetric in the
two scalar fields.

The inverse operation, bleaching, projects the dyed leg back to the
color-singlet state by replacing the generator
carried by $\bar H$ with the identity: $\blc_q: T^q \rightarrow 1$,
where we denote the bleaching operation on the Higgs leg with momentum $q$ by $\blc_q$.
This is a $U(1)$ decoupling operation and leads to
\begin{equation}
  \blc_q(f^{abq})=0 ,
  \qquad
  \blc_q(d^{abq})=\delta^{ab} .
  \label{eq:bleach_identities}
\end{equation}
Since bleaching acts only on color factors and preserves all kinematics,
applying it to the dyed amplitude gives the original FF:
\begin{equation}
  \blc_q\!\left[
    \amp_{n+1}(\{p_i\},q^{\bar H})
  \right]
  =
  \amp_{n+1}(\{p_i\},q^H)
    =
  \ff_n(\{p_i\}) .
  \label{eq:bleach_amp}
\end{equation}

The same dyeing and bleaching operations generalize directly to multi-operator cases.
Since the operator momenta are generally different with $q_i^2=m_i^2$,
one needs to introduce multiple Higgs fields, $H_i$, as well as their dyed versions, ${\bar H}_i$ associated with mass $m_i$.
The kinematic terms and the effective vertices are the same as in \eqref{eq:dyed_kinetic_terms} and \eqref{eq:dyed_operator_vertex}.

In the next two sections, we first consider the FFs with one operator insertion, i.e.~single-Higgs amplitudes, then we generalize to multi-Higgs cases.

\section{Single-operator case}

\begin{figure}[t]
	\centering
	\begin{tikzpicture}[
		line cap=round,
		line join=round,
		lab/.style={font=\scriptsize, inner sep=1pt},
		graphlab/.style={font=\scriptsize, inner sep=1pt},
		higgs/.style={
			draw=cyan!70!black,
			line width=1.55pt
		},
		scalarline/.style={draw=black, line width=0.55pt},
		gluonline/.style={
			draw=none,
			decoration={name=none},
			postaction={
				draw,
				line width=0.55pt,
				decoration={complete sines, amplitude=1.4mm, segment length=2mm},
				decorate=true
			}
		}
	]
	\coordinate (q1) at (-0.68,0);
	\coordinate (v1) at (0,0);
	\coordinate (j1) at (0.27,0.42);
	\coordinate (p11) at (0.58,0.92);
	\coordinate (p21) at (0.58,-0.92);
	\coordinate (g1) at (0.95,0);

	\coordinate (q2) at (2.10,0);
	\coordinate (v2) at (2.78,0);
	\coordinate (j2) at (3.05,-0.42);
	\coordinate (p12) at (3.36,0.92);
	\coordinate (p22) at (3.36,-0.92);
	\coordinate (g2) at (3.73,0);

	\coordinate (q3) at (4.95,0);
	\coordinate (j3) at (5.29,0);
	\coordinate (v3) at (5.63,0);
	\coordinate (p13) at (6.21,0.92);
	\coordinate (p23) at (6.21,-0.92);
	\coordinate (g3) at (5.29,0.92);

	\draw[higgs] (q1) -- (v1);
	\draw[scalarline] (v1) -- (j1) -- (p11);
	\draw[scalarline] (v1) -- (p21);
	\draw[gluonline] (j1) -- (g1);
	\draw[higgs] (q2) -- (v2);
	\draw[scalarline] (v2) -- (p12);
	\draw[scalarline] (v2) -- (j2) -- (p22);
	\draw[gluonline] (j2) -- (g2);
	\draw[gluonline] (j3) -- (g3);
	\draw[higgs] (q3) -- (j3) -- (v3);
	\draw[scalarline] (v3) -- (p13);
	\draw[scalarline] (v3) -- (p23);
	\node[lab, anchor=east, yshift=1.5pt] at (q1) {$q^{\bar H}$};
	\node[lab, anchor=south west] at (p11) {$1^\phi$};
	\node[lab, anchor=north west] at (p21) {$2^\phi$};
	\node[lab, anchor=west] at (g1) {$3^g$};
	\node[graphlab] at (0,-1.55) {(1)};

	\node[lab, anchor=east, yshift=1.5pt] at (q2) {$q^{\bar H}$};
	\node[lab, anchor=south west] at (p12) {$1^\phi$};
	\node[lab, anchor=north west] at (p22) {$2^\phi$};
	\node[lab, anchor=west] at (g2) {$3^g$};
	\node[graphlab] at (2.78,-1.55) {(2)};

	\node[lab, anchor=east, yshift=1.5pt] at (q3) {$q^{\bar H}$};
	\node[lab, anchor=south west] at (p13) {$1^\phi$};
	\node[lab, anchor=north west] at (p23) {$2^\phi$};
	\node[lab, anchor=south] at (g3) {$3^g$};
	\node[graphlab] at (5.56,-1.55) {(3)};
	\end{tikzpicture}
	\caption{Trivalent topologies of the dyed amplitude $\ampppgq$.
		Blue thick lines represent the dyed Higgs $\bar{H}$,
		black solid lines the scalars $\phi$,
		and wavy lines the gluons.}
	\label{fig:a4_cubic}
\end{figure}
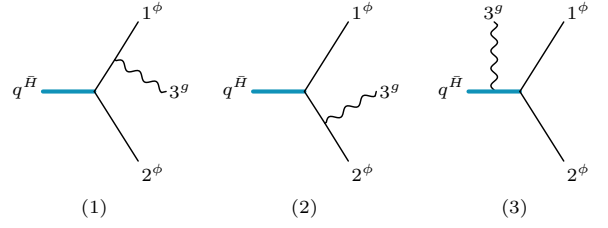

\noindent
The three-point FF $\fftabc$ provides an instructive example; at tree level it has two channels (as the first two graphs in FIG.~\ref{fig:a4_cubic}):
\begin{equation}
	\fftabc =
	\frac{C_{1}^{\Cf} N_{1}^{\Cf}}{s_{13}} +
	\frac{C_{2}^{\Cf} N_{2}^{\Cf}}{s_{23}}\,.
	\label{eq:f3_cubic_graphs}
\end{equation}
This example is discussed in detail in \cite{Lin:2021pne}, and the CK-dual numerators contain spurious poles.
Here we consider the corresponding dyed amplitude $\ampppgq$ which has a third channel through the new $g\bar{H}\bar{H}$ vertex (see FIG.~\ref{fig:a4_cubic}):
\begin{equation}
	\ampppgq =
	\frac{C_{1}^{\Ca} N_{1}^{\Ca}}{s_{13}} +
	\frac{C_{2}^{\Ca} N_{2}^{\Ca}}{s_{23}} +
	\frac{C_{3}^{\Ca} N_{3}^{\Ca}}{s_{12} - m^2}\,.
	\label{eq:a4_cubic_graphs}
\end{equation}
The color factors involve both $f$ and $d$ tensors:
\begin{gather}
	C_{1}^{\Ca} = f^{a_1 a_3 b} d^{b a_2 q}\,,
	\quad
	C_{2}^{\Ca} = f^{a_3 a_2 b} d^{b q a_1}\,,
	\notag \\
	C_{3}^{\Ca} = f^{q a_3 b} d^{b a_1 a_2}\, ,
	\label{eq:a4_cfs}
\end{gather}
which obey the generalized Jacobi relation:
\begin{equation}
	C_3^{\Ca} = -C_1^{\Ca} + C_2^{\Ca}\,.
	\label{eq:a4_color_relation}
\end{equation}
The Feynman-rule kinematic numerators,
\begin{gather}
	N_{1}^{\Ca} = -2 (p_1 \cdot \epsilon_3) = N_{1}^{\Cf} \,, \quad
	N_{2}^{\Ca} = 2 (p_2 \cdot \epsilon_3) = N_{2}^{\Cf}\,,
	\notag \\
	N_{3}^{\Ca} = 2 (p_1 + p_2) \cdot \epsilon_3\, ,
	\label{eq:a4_numerators_feyn}
\end{gather}
satisfy
\begin{equation}
	N_3^{\Ca} = - N_1^{\Ca} + N_2^{\Ca} \,,
\end{equation}
making CK duality manifest.

The double copy is performed by replacing color factors by numerators $C_i^{\Ca} \to N_i^{\Ca}$ in (\ref{eq:a4_cubic_graphs}), giving the gravity amplitude:
\begin{equation}
	\gapphq
	  = -4 \frac{(s_{13}(p_2 \cdot \epsilon_3) - s_{23}(p_1 \cdot \epsilon_3))^2}
	{s_{13} s_{23} (s_{12} - m^2)} \, ,
	\label{eq:m4_expression}
\end{equation}
where $h$ denotes the graviton.
This coincides with the double copy of $\ff_3$ \cite{Lin:2021pne}, satisfying both diffeomorphism invariance and factorization properties.

We now apply the bleaching operator $\blc_q$ to the constructed dyed amplitude.
The color factors are transformed as
\begin{gather}
	\blc(C_1^{\Ca}) = f^{a_1 a_3 b}\, \delta^{b a_2} = f^{a_1 a_3 a_2} = {\tilde C}_1^{\Cf}\,,
	\notag \\
	\blc(C_2^{\Ca}) = f^{a_3 a_2 b}\, \delta^{b a_1} = f^{a_1 a_3 a_2} = {\tilde C}_2^{\Cf}\,,
	\notag \\
	\blc(C_3^{\Ca}) = 0 \times d^{b a_1 a_2} = 0 \equiv {\tilde C}_3^{\Cf}\,.
	\label{eq:bleach_colors}
\end{gather}
Since $\blc_q$ acts only on color factors,
the bleached amplitude is
\begin{align}
	\blc_q(\amp_4)
	&= \frac{\blc_q(C_1^{\Ca}) N_1^{\Ca}}{s_{13}}
	+ \frac{\blc_q(C_2^{\Ca}) N_2^{\Ca}}{s_{23}}
	+ \frac{\blc_q(C_3^{\Ca}) N_3^{\Ca}}{s_{12} - m^2}
	\notag \\
	&= \frac{{\tilde C}_1^{\Cf} {\tilde N}_1^{\Cf}}{s_{13}}
	+ \frac{{\tilde C}_2^{\Cf} {\tilde N}_2^{\Cf}}{s_{23}}
	+ \frac{{\tilde C}_3^{\Cf} {\tilde N}_3^{\Cf}}{s_{12} - m^2}
	= \ff_3 \,,
	\label{eq:bleach_to_ff}
\end{align}
which is equal to the $\ff_3$ result in (\ref{eq:bleach_amp}).

Eq.\eqref{eq:bleach_to_ff} provides a novel CK-dual representation for the $\ff_3$ without spurious poles in the numerators.
Note that the third graph is only an \emph{auxiliary} graph, since ${\tilde C}_3^{\Cf}=0$,
and the massive pole $(s_{12} - m^2)$ does not contribute to the FF.
In contrast, in the previous construction based on \eqref{eq:f3_cubic_graphs} \cite{Lin:2021pne},
$N_1^{\Cf} = N_2^{\Cf}$ is forced by CK duality, and the spurious pole has to be introduced into the numerators.

We can also give a new interpretation for the hidden factorization of the FF.
Choosing $\{C^{\Ca}_2,\, C^{\Ca}_3\}$ as the Del Duca-Dixon-Maltoni (DDM) color basis~\cite{DelDuca:1999sg}, (\ref{eq:a4_cubic_graphs}) can be decomposed as
\begin{equation}
	\amp_4 = C^{\Ca}_2 \catracbq + C^{\Ca}_3 \catrcabq \,.
	\label{eq:a4_color_decompose}
\end{equation}
where $\ca_4$ denotes the color-ordered partial amplitudes.
Comparing (\ref{eq:a4_color_decompose}) with (\ref{eq:a4_cubic_graphs}) gives
\begin{equation}
	\vec{\ca}_4 = \Theta^{\amp} \cdot \vec{N}^{\amp},\
	\vec{\ca}_4 = \begin{bmatrix}
		\catracbq \\ \catrcabq
	\end{bmatrix},\
	\vec{N}^{\amp} = \begin{bmatrix}
		N_2^{\amp} \\ N_3^{\amp}
	\end{bmatrix}
\end{equation}
with the $\Theta^{\amp}$ matrix composed of propagators
\begin{equation}
	\Theta^{\amp} =
	\begin{bmatrix}
			\frac{1}{s_{13}} + \frac{1}{s_{23}} & -\frac{1}{s_{13}} \\
			-\frac{1}{s_{13}} & \frac{1}{s_{12} - m^2} + \frac{1}{s_{13}}
	\end{bmatrix}\,.
\end{equation}
The matrix $\Theta^{\amp}$ has rank $1$ due to kinematic constraints.
Its null space gives the BCJ relation
\begin{equation}
	-s_{23} \catracbq + (s_{12} - m^2) \catrcabq = 0 \,.
	\label{eq:a4_bcj_tr}
\end{equation}
Taking the residue of~(\ref{eq:a4_bcj_tr}) on the massive channel $s_{12} - m^2$
factorizes the second term:
\begin{align}
	&\lim_{s_{12} \to m^2}(s_{12}-m^2 )  \catrcabq
	\notag \\
	&= \ca_3(1^{\phi}, 2^{\phi}, -\mathbf{P}_{12}^{\bH})
	\times
	\ca_3(\mathbf{P}_{12}^{\bH}, q^{\bH}, 3^g)
\end{align}
where $\mathbf{P}_{12}=p_1+p_2$.
Moreover, $\catracbq$ receives contributions only from the first two topologies in FIG.~\ref{fig:a4_cubic}, and therefore equals the color-ordered FF
\begin{equation}
	\catracbq = \cftracb\,.
	\label{eq:a4_bia}
\end{equation}
Therefore, (\ref{eq:a4_bcj_tr}) reduces to
\begin{equation}
	\left. s_{23}\, \cftracb \right|_{s_{12} = m^2} = \cf_2(1^{\phi}, 2^{\phi})
	\ca_3(\mathbf{P}_{12}^{\bH}, q^{\bH}, 3^g) \,,
	\label{eq:a4_bcj_to_hidfac}
\end{equation}
where $\cf_2(1^\phi,2^\phi)=\ca_3(1^\phi,2^\phi,-\mathbf{P}_{12}^{\bH})=1$.
This matches precisely the hidden factorization relation of $\ff_3$  \cite{Lin:2021pne}.

This example illustrates the two central roles played by the dyeing framework.
First, dyeing embeds the FF into a fully colored amplitude, where the pole
that appears spurious in the FF construction is an ordinary massive
propagator of the dyed Higgs.
After bleaching, the corresponding color factor vanishes, but its
kinematic numerator remains part of the CK-dual structure and therefore
survives the double copy.
Second, the hidden factorization relations, which are obscure in the FF picture, are nothing but the BCJ relations of the dyed theory in the massive-channel on-shell limit.

\section{Multi-operator cases}

\noindent
We now extend the dyeing construction to FFs with multiple operator insertions
\begin{equation}
\ff_{n}^{[m]} = \int \prod_{i=1}^m d^{D} x_i\, e^{i q_i \cdot x_i}
\langle 1 \cdots n |\prod_{i=1}^m \mathcal{O}_i(x_i)| 0\rangle \,.
\end{equation}
Each operator insertion is represented by an adjoint massive Higgs particle $\bar H_i$ carrying momentum $q_i$ and mass $m_i^2=q_i^2$.
As in the previous section, we focus on the Yang--Mills-scalar theory with all operators $\mathcal O_i=\tr(\phi^2)$.

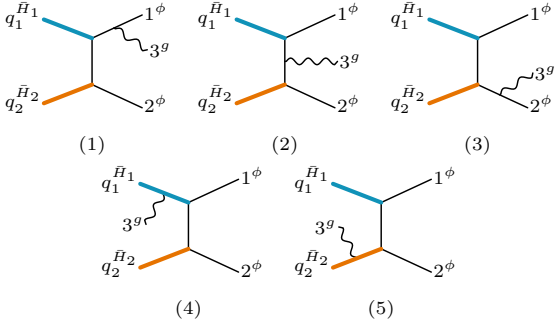
\begin{figure}[t]
	\centering
	\def\aFiveTopSkeleton{%
		\coordinate (T) at (0,0.36);
		\coordinate (B) at (0,-0.36);
		\coordinate (Qone) at (-0.75,0.65);
		\coordinate (Qtwo) at (-0.75,-0.65);
		\coordinate (Pone) at (0.78,0.72);
		\coordinate (Ptwo) at (0.78,-0.72);
		\draw[scalarline] (T) -- (B);
		\draw[scalarline] (T) -- (Pone);
		\draw[scalarline] (B) -- (Ptwo);
	}%
	\def\aFiveTopHiggs{%
		\draw[higgsone] (Qone) -- (T);
		\draw[higgstwo] (Qtwo) -- (B);
	}%
	\def\aFiveTopLabels{%
		\node[lab, anchor=east, yshift=1.5pt] at (Qone) {$q_1^{\bar H_1}$};
		\node[lab, anchor=east, yshift=1.5pt] at (Qtwo) {$q_2^{\bar H_2}$};
		\node[lab, anchor=west, yshift=1.5pt] at (Pone) {$1^\phi$};
		\node[lab, anchor=west, yshift=1.5pt] at (Ptwo) {$2^\phi$};
	}%
	\begin{tikzpicture}[
		x=0.85cm,
		y=0.85cm,
		line cap=round,
		line join=round,
		lab/.style={font=\scriptsize, inner sep=0.6pt},
		graphlab/.style={font=\scriptsize, inner sep=1pt},
		higgsone/.style={
			draw=cyan!70!black,
			line width=1.55pt
		},
		higgstwo/.style={
			draw=orange!90!black,
			line width=1.55pt
		},
		scalarline/.style={draw=black, line width=0.55pt},
		gluonline/.style={
			draw=none,
			decoration={name=none},
			postaction={
				draw,
				line width=0.55pt,
				decoration={complete sines, amplitude=1.4mm, segment length=2mm},
				decorate=true
			}
		}
	]
	\begin{scope}[shift={(0,0)}]
		\aFiveTopSkeleton
		\draw[gluonline] (0.34,0.52) -- (0.86,0.18);
		\aFiveTopHiggs
		\aFiveTopLabels
		\node[lab, anchor=west] at (0.86,0.18) {$3^g$};
		\node[graphlab] at (0,-1.30) {$(1)$};
	\end{scope}
	\begin{scope}[shift={(3.00,0)}]
		\aFiveTopSkeleton
		\draw[gluonline] (0,0) -- (0.82,0);
		\aFiveTopHiggs
		\aFiveTopLabels
		\node[lab, anchor=west] at (0.82,0) {$3^g$};
		\node[graphlab] at (0,-1.30) {$(2)$};
	\end{scope}
	\begin{scope}[shift={(6.00,0)}]
		\aFiveTopSkeleton
		\draw[gluonline] (0.34,-0.52) -- (0.86,-0.18);
		\aFiveTopHiggs
		\aFiveTopLabels
		\node[lab, anchor=west] at (0.86,-0.18) {$3^g$};
		\node[graphlab] at (0,-1.30) {$(3)$};
	\end{scope}
	\begin{scope}[shift={(1.50,-2.55)}]
		\aFiveTopSkeleton
		\draw[gluonline] (-0.38,0.51) -- (-0.66,0.02);
		\aFiveTopHiggs
		\aFiveTopLabels
		\node[lab, anchor=east] at (-0.66,0.02) {$3^g$};
		\node[graphlab] at (0,-1.30) {$(4)$};
	\end{scope}
	\begin{scope}[shift={(4.50,-2.55)}]
		\aFiveTopSkeleton
		\draw[gluonline] (-0.38,-0.51) -- (-0.66,-0.02);
		\aFiveTopHiggs
		\aFiveTopLabels
		\node[lab, anchor=east] at (-0.66,-0.02) {$3^g$};
		\node[graphlab] at (0,-1.30) {$(5)$};
	\end{scope}
	\end{tikzpicture}
	\caption{
		Trivalent topologies for the ordering sector $\ampTrsA$ of $\dampPPGHH$.
		Blue and orange thick lines denote the dyed Higgs bosons $\bar{H}_1$ and $\bar{H}_2$.
		The sector with order $[12q_1q_2]$ follows from $q_1\leftrightarrow q_2$.}
	\label{fig:a5_trivalent}
\end{figure}

The simplest nontrivial case is the five-point dyed amplitude $\dampPPGHH$, which contains two distinct dyed Higgs bosons $\bar{H}_1$ and $\bar{H}_2$ and corresponds to a generalized two-operator-insertion FF with off-shell momenta $q_1$, $q_2$.
This amplitude admits the decomposition
\begin{equation}
	\dampPPGHH =
	\ampTrsA + (q_1 \leftrightarrow q_2) \,.
	\label{eq:a5_decompose}
\end{equation}
Here $\trsA$ denotes the cyclic ordering of the color labels carried by the external scalar lines, namely the two $\phi$ legs and the dyed Higgs legs.
We call $\ampTrsA$ a \emph{scalar-ordered} amplitude, obtained by keeping the scalar ordering fixed and summing over all possible gluon insertions: this gives the five cubic graphs in FIG.~\ref{fig:a5_trivalent}
\begin{equation}
  \ampTrsA = \sum_{i = 1}^{5} \frac{C_{\clsA|i}
  N_{\clsA|i}}{D_{\clsA|i}}\,, \quad I=  \trsA \,.
\end{equation}
The other sector is obtained by exchanging the $q_1$ and $q_2$ legs.

This decomposition follows from the color structure.
The color-Jacobi relations do not connect topologies with different scalar orderings.
They close within each fixed scalar-ordering sector.
For example,
\begin{equation}
	\begin{tikzpicture}[
		x=0.58cm,
		y=0.58cm,
		baseline={(current bounding box.center)},
		line cap=round,
		line join=round,
		lab/.style={font=\scriptsize, inner sep=0.6pt},
		higgsone/.style={
			draw=cyan!70!black,
			line width=1.55pt
		},
		higgstwo/.style={
			draw=orange!90!black,
			line width=1.55pt
		},
		higgsthree/.style={
			draw=green!70!black,
			line width=1.55pt
		},
		scalarline/.style={draw=black, line width=0.55pt},
		gluonline/.style={
			draw=none,
			decoration={name=none},
			postaction={
				draw,
				line width=0.55pt,
				decoration={complete sines, amplitude=1.4mm, segment length=2mm},
				decorate=true
			}
		},
		vtx/.style={circle, fill=white, draw=black, line width=0.45pt, inner sep=0pt, minimum size=3.1pt}
	]
	\def\ddmLines{%
		\coordinate (L) at (-0.75,0);
		\coordinate (R) at (2.75,0);
		\coordinate (Qone) at (0,0);
		\coordinate (Qtwo) at (1.05,0);
		\coordinate (Qthree) at (2.10,0);
		\draw[scalarline] (L) -- (R);
		\draw[higgsone] (Qone) -- (0,1.05);
		\draw[higgstwo] (Qtwo) -- (1.05,1.05);
		\draw[higgsthree] (Qthree) -- (2.10,1.05);
	}%
	\def\ddmVertices{%
		\node[vtx] at (Qone) {};
		\node[vtx] at (Qtwo) {};
		\node[vtx] at (Qthree) {};
	}%
	\def\ddmLabels{%
		\node[lab, anchor=east] at (L) {$\phi$};
		\node[lab, anchor=west] at (R) {$\phi$};
		\node[lab, anchor=south] at (0,1.05) {$\bar H_1$};
		\node[lab, anchor=south] at (1.05,1.05) {$\bar H_2$};
		\node[lab, anchor=south] at (2.10,1.05) {$\bar H_3$};
	}%
	\begin{scope}[shift={(0,0)}]
		\ddmLines
		\draw[gluonline] (1.05,0.55) to[out=0,in=-110] (1.55,1.05);
		\ddmVertices
		\node[vtx] at (1.05,0.55) {};
		\ddmLabels
		\node[lab, anchor=south] at (1.55,1.05) {$g$};
	\end{scope}
	\node[font=\normalsize] at (3.55,0.47) {$=$};
	\begin{scope}[shift={(5.10,0)}]
		\ddmLines
		\draw[gluonline] (1.58,0) -- (1.58,1.05);
		\ddmVertices
		\node[vtx] at (1.58,0) {};
		\ddmLabels
		\node[lab, anchor=south] at (1.58,1.05) {$g$};
	\end{scope}
	\node[font=\normalsize] at (8.58,0.47) {$-$};
	\begin{scope}[shift={(10.05,0)}]
		\ddmLines
		\draw[gluonline] (0.52,0) -- (0.52,1.05);
		\ddmVertices
		\node[vtx] at (0.52,0) {};
		\ddmLabels
		\node[lab, anchor=south] at (0.52,1.05) {$g$};
	\end{scope}
	\end{tikzpicture}\,.
	\label{eq:ddm_closure}
\end{equation}
Consequently, the full amplitude decomposes into disjoint gauge-invariant
sectors labeled by scalar orderings.

The five color factors in  FIG.~\ref{fig:a5_trivalent} satisfy the following Jacobi relations:
\begin{equation}
  C_{\clsA|4} = C_{\clsA|1} - C_{\clsA|2}, \quad C_{\clsA|5} =
  -C_{\clsA|2} + C_{\clsA|3}\, .
  \label{eq:a5_color_relations}
\end{equation}
The CK-dual numerators can be constructed, giving a consistent gravity amplitude after double copy:
\begin{equation}\label{eq:A5toM5scalarorder}
	\mathscr{A}^{[12q_2q_1]}_5 \xrightarrow{\text{double copy}} \mathscr{M}^{[12q_2q_1]}_5\,.
\end{equation}

To understand this property, the key is to consider the propagator matrix.
Taking the $\{C_{\clsA|i}, i=1,2,3\}$ as the DDM color basis, the propagator matrix of the scalar-ordered amplitude is
\begin{gather}
  \ppmAppghhI =
  \begin{bmatrix}
    \frac{1}{D_{\clsA|1}} + \frac{1}{D_{\clsA|4}} &
    -\frac{1}{D_{\clsA|4}} & 0 \\
    -\frac{1}{D_{\clsA|4}} & \frac{1}{D_{\clsA|2}} + \frac{1}{D_{\clsA|4}} +
    \frac{1}{D_{\clsA|5}} & -\frac{1}{D_{\clsA|5}} \\
    0 & -\frac{1}{D_{\clsA|5}} & \frac{1}{D_{\clsA|3}} + \frac{1}{D_{\clsA|5}}
  \end{bmatrix} .
\end{gather}
Crucially, $\ppmAppghhI$ is singular and has rank 2.
This leads to two important properties.

\newcommand{\kltAppghhI}{\klt^{\trsA}_{2 \times 2}}

First, the KLT kernel is given by a $2\times 2$ matrix
\begin{equation}
  \kltAppghhI =
  \begin{bmatrix}
    \frac{1}{D_{\clsA|1}} + \frac{1}{D_{\clsA|4}} & -\frac{1}{D_{\clsA|4}} \\
    -\frac{1}{D_{\clsA|4}} & \frac{1}{D_{\clsA|2}} + \frac{1}{D_{\clsA|4}} +
    \frac{1}{D_{\clsA|5}}
  \end{bmatrix}^{-1}\, ,
\end{equation}
which contains only physical poles. In contrast, in the non-dyeing picture, because only the first three topologies in FIG.~\ref{fig:a5_trivalent} contribute, the resulting propagator matrix and KLT kernel involve highly complicated unphysical poles that invalidate the double copy.

Second,
the null space of $\ppmAppghhI$ yields a BCJ relation among amplitudes of the scalar order $\trsA$:
\begin{align}
	 & - (s_{3q_1} - m_1^2)
	\campnh{2}{5}(3^g,1^\phi,2^\phi,q_2^{\bH_2},q_1^{\bH_1}) \notag \\
	 & + (s_{3q_2} - m_2^2)
	\campnh{2}{5}(1^\phi,2^\phi,3^g,q_2^{\bH_2},q_1^{\bH_1}) \notag \\
	 & - (s_{3q_1} + s_{13} - m_1^2)
	\campnh{2}{5}(1^\phi,3^g,2^\phi,q_2^{\bH_2},q_1^{\bH_1}) = 0 \,.
	\label{eq:a5_bcj_tr}
\end{align}
Imposing the on-shell conditions on the two massive poles
$(s_{3q_1} - m_1^2)$ and $(s_{3q_2} - m_2^2)$,
and identifying the corresponding color-ordered amplitudes with the
color-ordered FFs, we obtain the factorization relation
for the two-operator FF:
\begin{align}
	& \quad\ \left.s_{13}
		\cffnh{2}{3}(1^\phi, 3^g, 2^\phi, q_2^{H_2}, q_1^{H_1})
	\right|_{s_{3q_2} = m_{2}^2}^{s_{3q_1} = m_{1}^2}
	\\
	= & -\left[
		\ca_3(q_1^{\bH_1}, 3^g, -\mathbf{P}_{3q_1}^{\bH_1})
		\cffnh{2}{2}(1^\phi, 2^\phi, q_2^{H_2}, \mathbf{P}_{3q_1}^{H_1})
	\right]_{s_{3q_2} = m_{2}^2}
	\notag \\
	&+\left[
		\ca_3(-\mathbf{P}_{3q_2}^{\bH_2}, 3^g, q_2^{\bH_2})
		\cffnh{2}{2}(1^\phi, 2^\phi, \mathbf{P}_{3q_2}^{H_2}, q_1^{H_1})
	\right]_{s_{3q_1} = m_{1}^2}\,,
	\notag
	\label{eq:f3_hidden_fac}
\end{align}
which generalizes the hidden factorization relations for the single-operator FFs.

\begin{figure}[t]
  \centering
  \subfigure[Full-color]{%
    \begin{minipage}{.31\linewidth}
      \centering
      \includegraphics[width=.74\linewidth]{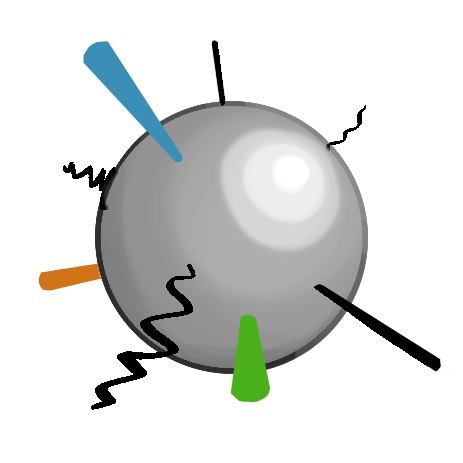}
    \end{minipage}}
  \hfill
  \subfigure[Color-ordered]{%
    \begin{minipage}{.31\linewidth}
      \centering
      \begin{tikzpicture}[
        x=1cm,
        y=1cm,
        line cap=round,
        line join=round,
        higgsone/.style={draw=cyan!70!black, line width=1.55pt},
        higgstwo/.style={draw=orange!90!black, line width=1.55pt},
        higgsthree/.style={draw=letterGreen, line width=1.55pt},
        scalarline/.style={draw=black, line width=0.55pt},
        gluonline/.style={
          draw=none,
          decoration={name=none},
          postaction={
            draw,
            line width=0.55pt,
            decoration={complete sines, amplitude=1.4mm, segment length=2mm},
            decorate=true
          }
        }
      ]
        \def\ampRadius{0.52}
        \def\legRadius{0.96}
        \draw[higgsone] (90:\ampRadius) -- (90:\legRadius);
        \draw[gluonline] (45:\ampRadius) -- (45:\legRadius);
        \draw[scalarline] (0:\ampRadius) -- (0:\legRadius);
        \draw[higgsthree] (-45:\ampRadius) -- (-45:\legRadius);
        \draw[gluonline] (-90:\ampRadius) -- (-90:\legRadius);
        \draw[higgstwo] (-135:\ampRadius) -- (-135:\legRadius);
        \draw[scalarline] (180:\ampRadius) -- (180:\legRadius);
        \draw[gluonline] (135:\ampRadius) -- (135:\legRadius);
        \filldraw[fill=gray!40, draw=black, line width=0.45pt] (0,0) circle (\ampRadius);
      \end{tikzpicture}
    \end{minipage}}
  \hfill
  \subfigure[Scalar-ordered]{%
    \begin{minipage}{.31\linewidth}
      \centering
      \includegraphics[width=.74\linewidth]{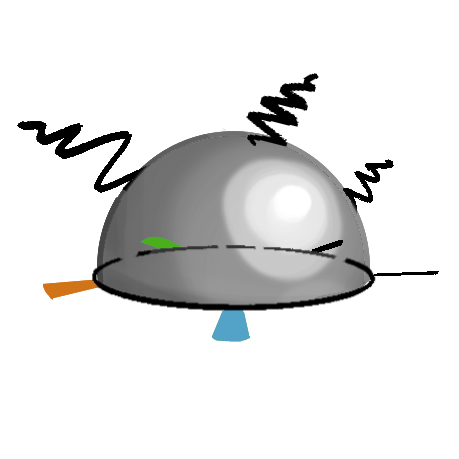}
    \end{minipage}}
	\caption{
		Graphic representation of full-color, color-ordered, and scalar-ordered amplitudes.
		(a)~Full-color: all states on a sphere with no ordering.
		(b)~Color-ordered: a planar disk with all particles ordered on the boundary.
		(c)~Scalar-ordered: scalars fixed on the hemisphere boundary with a prescribed order $[{\clrBlue \bar{H}_1}{\clrOrange \bar{H}_2}{\clrLetterGreen \bar{H}_3}\phi\phi]$, while gluons remain unordered on the surface.
	}
	\label{fig:scalar_decompose}
\end{figure}
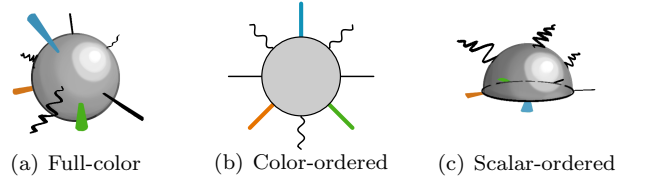

\begin{figure}[t]
	\centering
	\[
		\vcenter{\hbox{\raisebox{0.12cm}{\includegraphics[width=0.24\linewidth]{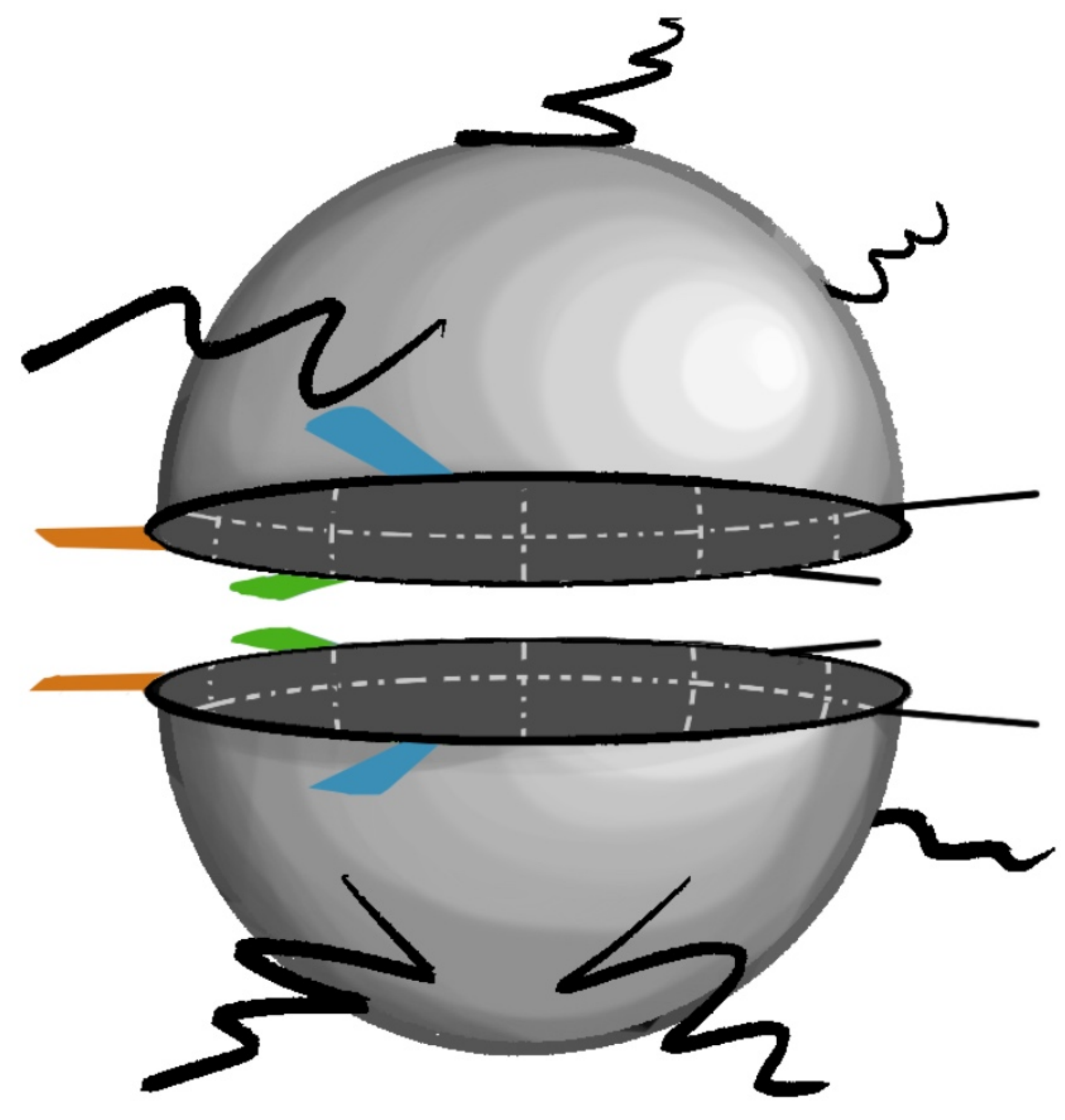}}}}
		\hspace{0.55cm}
		\xrightarrow{\text{ double copy }}
		\hspace{0.55cm}
		\vcenter{\hbox{\includegraphics[width=0.24\linewidth]{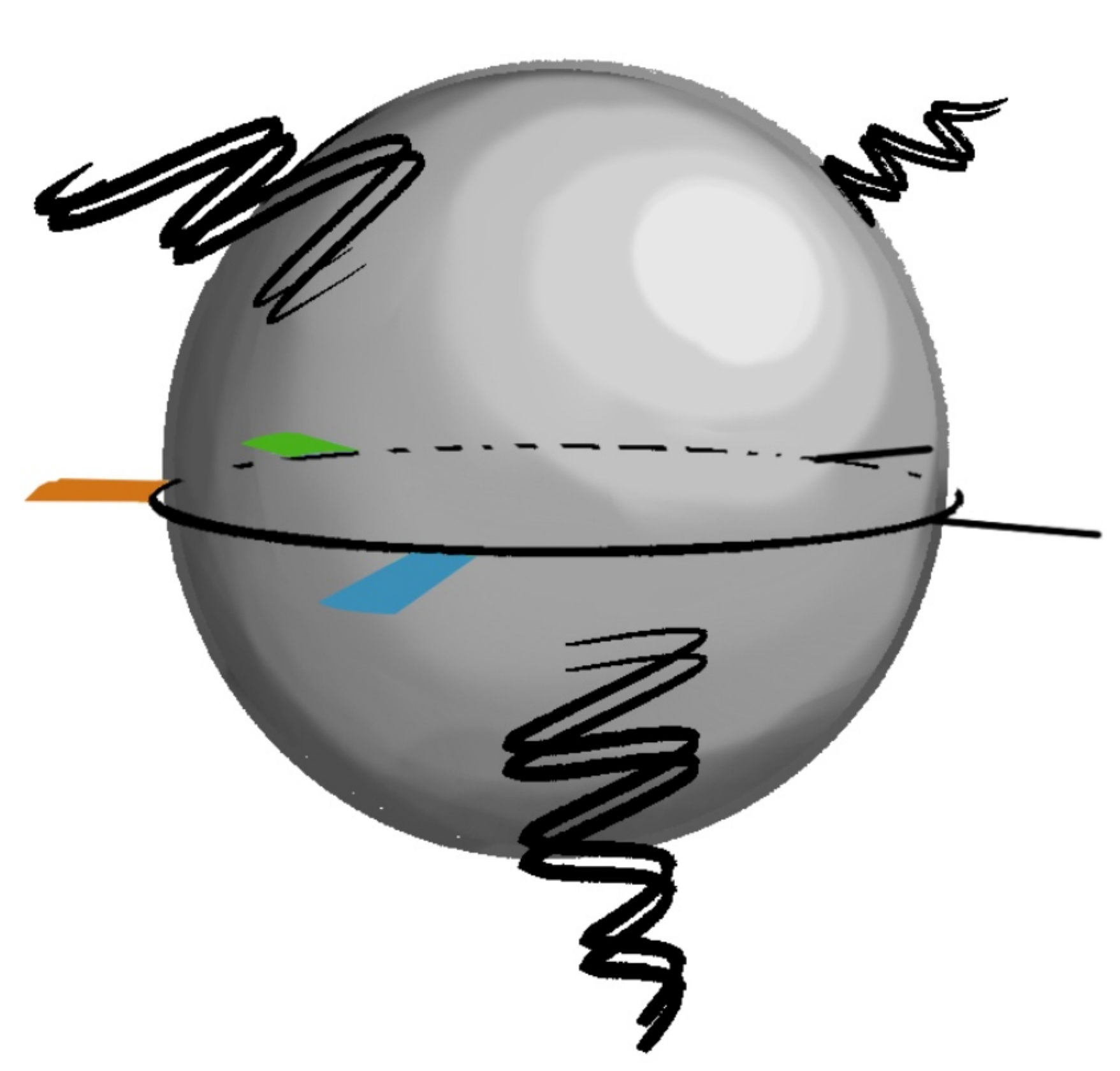}}}
	\]
	\caption{Double copy of scalar-ordered gauge amplitudes.
		Gravitons are denoted by double wavy lines.
		A nonzero gravity contribution arises only from compatible scalar orderings in the two gauge copies.}
	\label{fig:scalar_ordered_dc}
\end{figure}

We further comment on scalar ordering and the double-copy results.
An illustration of full-color, color-ordered, and scalar-ordered gauge-theory amplitudes is presented in FIG.~\ref{fig:scalar_decompose}.
In particular, the scalar-ordered amplitudes can be viewed as a hemisphere where all scalars are ordered on the boundary.
At the gravity level, all states are colorless, so there is a priori no
color ordering.
However, the KLT construction preserves the scalar-ordering.
For the above five-point example, the gravity amplitude is constructed as
\begin{equation}
	\ganh{2}{5}=\vec{\ca}^{\,T}\cdot\klt\cdot\vec{\ca}\, ,
\end{equation}
where the KLT kernel $\klt$ is obtained by inverting the full-rank part of the propagator matrix,
and $\vec{\ca}$ is composed of color-ordered amplitudes.
Since the propagator matrix is block-diagonal, $\klt$ has the same block structure:
matrix elements of scalar orderings are decoupled in the double copy.
Thus only double copies of compatible scalar-ordered gauge amplitudes yield nonzero contributions.
Each scalar-ordered gauge sector gives an independent, diffeomorphism-invariant scalar-ordered gravity sector as in \eqref{eq:A5toM5scalarorder},
\newcommand{\GAppghh}{\ganh{2}{5}}
and the full gravity result is given by
\begin{equation}
	\GAppghh =
	\mathscr{M}^{[12q_2q_1]}_5 + (q_1 \leftrightarrow q_2) \,.
	\label{eq:M5_decompose}
\end{equation}
This double-copy and decomposition picture is illustrated in FIG.~\ref{fig:scalar_ordered_dc},
where two hemispheres are glued and the scalars are ordered on the equator.

The above discussion generalizes straightforwardly to higher-point cases with multi-Higgs particles.
Sample counts up to eight points are given in Table~\ref{tab:multi_higgs_compact}.
The table shows the following patterns.
For $\ampnh{n_h}{n}$, the $n$-point dyed amplitude with $n_h$ Higgs legs and $n_g=n-n_h-2$ gluons, there are $n_h!$ scalar orderings.
Each scalar-ordered propagator matrix has size and rank
\begin{equation}
	\#\text{SIZE} = \frac{(n-2)!}{n_h!}\,,
	\quad
	\#\text{RANK} = \frac{(n-3)!}{(n_h-1)!}\,.
	\label{eq:counting_general}
\end{equation}
The full propagator matrix therefore has rank $n_h\times (n-3)!$.
The rank-to-size ratio is
\begin{equation}
	\frac{\#\text{RANK}}{\#\text{SIZE}}
	=
	\frac{n_h}{n-2}
	=
	\frac{n-n_g-2}{n-2}\,.
\end{equation}
This ratio measures the degree of BCJ redundancy.
It approaches one when Higgs legs dominate, and the redundancy is reduced.
When gluons dominate, it approaches zero, as in pure-gluon amplitudes.

\begin{table}[t]
	\centering
	\begin{tabular}{c|c|c|c}
	\hline
		$n$ & $n_h$ & block size & block rank \\
		\hline
		5   & 2     & 3          & 2          \\
		6   & 2     & 12         & 6          \\
		7   & 3     & 20         & 12         \\
		8   & 4     & 30         & 20         \\ \hline
	\end{tabular}
		\caption{
		Counting data for each scalar-ordered block of $\ampnh{n_h}{n}$.
		``block size'' and ``block rank'' are the size and rank of each scalar-ordered propagator matrix.
		}
	\label{tab:multi_higgs_compact}
\end{table}

\section{Discussion}

\noindent
Our dyeing mechanism applies to the amplitudes involving color-singlet external particles, where the double copy faces challenges similar to those of FFs.
Note that the dyed Higgs is distinct from the usual colored Higgs fields in spontaneously broken gauge theories,
whose double copy has been studied in~\cite{Chiodaroli:2015rdg,Naculich:2015coa}.
Here, the dyed Higgs couples through $d$-symmetric color tensors, and
the dyed amplitudes constitute a new class of amplitudes.
We note that CK-dual constructions involving symmetric and mixed $d$-$f$ color structures have also been studied in the EFT double-copy context~\cite{Carrasco:2021ptp,Carrasco:2022jxn,Carrasco:2026hxf}.

Below we comment on two immediate generalizations:
\begin{enumerate}
\item
\emph{General operators.}---
For higher-dimensional operators $\text{tr}(\phi^m)$, the dyed vertex involves the fully symmetric tensor $d^{a_1\cdots a_{m+1}}$.
For a fermion bilinear $\bar\psi\psi$, the dyeing procedure promotes the operator to a Yukawa-like interaction $H^a \bar\psi T^a \psi$.
For a vector operator $\bar\psi\gamma^\mu\psi$, the mechanism introduces a dyed vector field through $V_\mu^a (\bar\psi \gamma^\mu T^a \psi)$,
which, upon double copy, becomes a spin-2 field coupled to the stress-energy tensor as $G_{\mu\nu}T^{\mu\nu}$. The construction proceeds in a similar manner as presented in the main text.

\item
\emph{Loop level.}---
In previous studies of CK-duality for loop-level FFs \cite{Boels:2012ew,Yang:2016ear,Lin:2021kht,Li:2024wzj}, CK relations on internal lines
directly connected to the operator were not imposed. Imposing full CK duality in these cases often introduces unphysical poles
 that are inconsistent with unitarity.
The dyed theory  offers a promising resolution to this issue: the operator propagates inside the loop as a massive internal state, supplying the additional topologies and CK numerators required to restore consistency.
Upon bleaching, these additional contributions have vanishing color factors, yielding FF integrands that satisfy unitarity checks and whose kinematic numerators can be double-copied to gravity integrands.
Preliminary studies suggest that this approach is indeed viable.
\end{enumerate}
A more comprehensive discussion of these results, including higher-point examples and extensions to other operators and loop levels, will be presented in a forthcoming companion paper.

\vskip .3cm

{\it Acknowledgments.}
We would like to thank Zhiming Cai, Zeyu Li and Guanda Lin for discussions.
This work is supported by the National Natural Science Foundation of
China (Grants No.~12425504, 12447101, 12247103) and by the Chinese
Academy of Sciences (Grant No.~YSBR-101).
We also acknowledge the support of the HPC Cluster of ITP-CAS.

\end{document}